\documentclass[12pt]{iopart}

\usepackage{iopams}
\usepackage{braket}
\usepackage{amsfonts}
\usepackage{color}
\usepackage[dvipsnames]{xcolor}
\usepackage{amssymb}
\usepackage{mathrsfs}
\usepackage{graphicx}
\usepackage{lmodern}
\usepackage{lineno}
\usepackage{hyperref}
\usepackage[normalem]{ulem}
\usepackage{soul}

\newcommand{\beq}{\begin{eqnarray}}
\newcommand{\eeq}{\end{eqnarray}}

\newcommand{\csch}{\mathrm{csch}}

\newcommand{\Exp}{\mathrm{Exp}_{\star}}
\newcommand{\Expg}{\mathrm{Exp}_{\star_{\gamma}}}

\def\keywords#1{\vspace{10pt}
     \begin{indented}
     \item[]\rm Keywords: #1\par
     \end{indented}}

\begin{document}



\title{The Feynman-Kac formula in deformation quantization}
\author{Jasel Berra--Montiel$^{1,2}$, Hugo Garc\'ia--Compe\'an$^{3}$ and Alberto Molgado$^{1}$}

\address{$^{1}$ Facultad de Ciencias, Universidad Aut\'onoma de San Luis 
Potos\'{\i} \\
Campus Pedregal, Av. Parque Chapultepec 1610, Col. Privadas del Pedregal, San
Luis Potos\'{\i}, SLP, 78217, Mexico}
\address{$^2$ Dipartimento di Fisica ``Ettore Pancini", Universit\'a degli studi di Napoli ``Federico II", Complesso Univ. Monte S. Angelo, I-80126 Napoli, Italy}
\address{$^3$Departamento de F\'isica, Centro de Investigaci\'on y de Estuios Avanzados del IPN \\
P.0. Box 14-740, Ciudad de M\'exico, 07000, Mexico}

\eads{\mailto{\textcolor{blue}{jasel.berra@uaslp.mx}},\ 
\mailto{\textcolor{blue}{hugo.compean@cinvestav.mx}}\
\mailto{\textcolor{blue}{alberto.molgado@uaslp.mx}} 
}


\begin{abstract}
We introduce the Feynman-Kac formula within the deformation quantization program. Constructing on previous work it is shown that, upon a Wick rotation, the ground state energy of any prescribed physical system can be obtained from the asymptotic limit of the phase space integration of the star exponential of the Hamiltonian operator. Some examples of this correspondence are provided showing a novel and efficient way of computing the ground state energy for some physical models.
\end{abstract}



\section{Introduction}

The modern concept of deformation quantization was first introduced in the seminal work by Bayen, Fronsdal, Flato, Lichnerowicz and Sternheimer~\cite{Bayen} as a general approach to pass from a classical mechanical system to its corresponding quantum system. The main feature within this quantization scheme lies on the algebraic structure of the observable algebra. In contrast to the standard formulation, where observables are represented by self-adjoint operators on a Hilbert space, deformation quantization preserves the same vector space of smooth functions defined on a Poisson manifold, corresponding to the classical phase space. However, the usual commutative pointwise product of functions is replaced by a noncommutative and associative product, known as the star product, where Planck's constant acts as a deformation parameter.

Although the deformation quantization framework leads to equivalent results as those obtained through the operator formalism of quantum mechanics, the former approach offers several conceptual advantages. The first one focuses on interpreting classical physics as a contraction of quantum theory in the limit $\hbar\to 0$. Here, the concept of contraction is understood in the sense of the Segal-Wigner-Inonu contraction of algebraic structures~\cite{Segal}, in which group theoretical techniques are employed to derive a ``contracted" or simplified group from a more complex one by taking a limit on certain parameters.  A well-known example is the derivation of the Galilean group from the Poincar\'e group in the limit as the speed of light $c\to\infty$, representing the transition from the relativistic to the non-relativistic framework. In this manner, a deformation can be interpreted as the inverse of a contraction, in the sense that by determining all groups that are close to the Galilei group, one is allowed to recover the Poincar\'e  group. This implies that relativistic theory can be seen as a deformation of non-relativistic physics. The second advantage lies in the generality of the deformation quantization formalism which only requires the existence of a symplectic structure or a Poisson algebra of classical observables for its implementation. The main challenge, however, is in establishing the existence of star products and classifying their possible forms. Nevertheless, the underlying conceptual framework and physical interpretation of observables are completely well established from the outset. \cite{Waldmann}. 

Within the framework of deformation quantization, the determination of spectral properties and the construction of the projectors associated with a given observable are achieved through the use of the star exponential. This particular function, defined on the phase space, is entirely formulated in terms of the star product and provides a topological completion of the Weyl algebra. Moreover, it also plays a crucial role in governing the evolution of the Wigner function as described by Moyal's equation \cite{Generating}. 

On the other hand, the Feynman-Kac formula offers a powerful tool for determining the ground state energy of a quantum system by linking the asymptotic behavior of the propagator to its lowest energy eigenvalue. Specifically, the ground state energy is expressed as the expectation value of an exponential function associated to a stochastic process \cite{Glimm}. Bearing this in mind, and considering recent connections between the star exponential function and propagators in the path integral formalism \cite{starexpo}, \cite{Sharan}, the main aim of the present paper is to establish the Feynman-Kac formula within the framework of deformation quantization.   

The paper is organized as follows. In Section \ref{sec:DQ} we briefly review the deformation quantization program and the Moyal product. In Section \ref{sec:FK} we explore the relation between the asymptotic behavior of the star exponential function and the Feynman-Kac formula. We also include in this section some simple examples to illustrate our construction. Finally, we include some concluding remarks and perspectives in Section \ref{sec:conclu}.    

\section{Deformation quantization and Moyal product}
\label{sec:DQ}

The origins of the deformation quantization formalism can be traced back to the Weyl's (Weyl-Wigner) quantization map \cite{Wigner}, which establishes a correspondence between any classical observable, $f(\mathbf{x},\mathbf{p})\in L^{2}(\mathbb{R}^{2n})$ defined on the phase space $\Gamma=\mathbb{R}^{2n}$, and an operator (the quantum observable) acting on the Hilbert space $\mathcal{H}=L^{2}(\mathbb{R}^{n})$, through the relation
\begin{equation} \label{Weyl}
\mathcal{Q}_{W}(f)=\frac{1}{(2\pi)^{n}}\int_{\mathbb{R}^{2n}}\widetilde{f}(\mathbf{a},\mathbf{b})e^{i(\mathbf{a}\cdot\mathbf{\widehat{X}}+\mathbf{b}\cdot\mathbf{\widehat{P}})}d\mathbf{a}\,d\mathbf{b}   \,,
\end{equation}
where $\widetilde{f}(\mathbf{a},\mathbf{b})$ denotes the Fourier transform on $\mathbb{R}^{2n}$, while the components of the operators $\widehat{\mathbf{X}}$, $\widehat{\mathbf{P}}$ satisfy the canonical commutation relations $[\widehat{X}_{i},\widehat{P}_{j}]=i\hbar\delta_{ij}$, with $i,j=1,\ldots, n$. Strictly  speaking, the integral is evaluated in the weak operator topology and may not be absolutely convergent \cite{Takhtajan} (see also \cite{Dias}, and references therein). Since the Weyl's transform corresponds to the integral of an operator, we can compute its integral kernel as
\begin{equation}
\kappa_{f}(\mathbf{x},\mathbf{x}')=\braket{\mathbf{x}|\mathcal{Q}_{W}(f)|\mathbf{x}'}.
\end{equation}
By means of the Baker-Campbell-Hausdorff formula, we obtain
\begin{equation}
\kappa_{f}(\mathbf{x},\mathbf{x}')=\frac{1}{(2\pi\hbar)^{n}}\int_{\mathbb{R}^{n}}f\left(\frac{\mathbf{x}+\mathbf{x}'}{2},\mathbf{p} \right)e^{i\mathbf{p}\cdot(\mathbf{x-\mathbf{x'}})/\hbar}d\mathbf{p}. 
\end{equation}
This integral kernel also satisfies the relation
\begin{equation}
\int_{\mathbb{R}^{2n}}|\kappa_{f}(\mathbf{x},\mathbf{x}')|^{2}d\mathbf{x}d\mathbf{x}'=\frac{1}{(2\pi\hbar)^{n}}\int_{\mathbb{R}^{2n}}|f(\mathbf{x},\mathbf{p})|^{2}d\mathbf{x}d\mathbf{p},
\end{equation}
which indicates that the kernel defines a Hilbert-Schmidt operator. Specifically, it maps functions from $L^{2}(\mathbb{R}^{n})$ into functions in $L^{2}(\mathbb{R}^{n})$  as 
\begin{equation}
\mathcal{Q}_{W}(f)\psi(\mathbf{x})=\int_{\mathbb{R}^{n}}\kappa_{f}(\mathbf{x},\mathbf{x}')\psi(\mathbf{x'})d\mathbf{x}',
\end{equation}
for any $\psi\in L(\mathbb{R}^{n})$. Moreover, this operator possesses a well defined (but possibly infinite) trace \cite{Reed}. We conclude that the Weyl transform (\ref{Weyl})  can be regarded as a mapping from the set of square-integrable functions on the phase space  $L^{2}(\mathbb{R}^{2n})$ to the space of Hilbert–Schmidt operators, $\mathrm{HS}(L^2(\mathbb{R}^n))$, acting on the Hilbert space $L^2(\mathbb{R}^n)$. In particular, it can be demonstrated that the Weyl's transform $\mathcal{Q}_{W}(f)$  corresponds to a self-adjoint operator whenever  $f$ is a real-valued function \cite{Curtright}.
A unique function on phase space can be determined by applying the inverse of Weyl's map $\mathcal{Q}_{W}^{-1}:\mathrm{HS}(L^2(\mathbb{R}^n))\to L^{2}(\mathbb{R}^{2n})$, which is defined as follows
\begin{equation}\label{inverseWeyl}
\mathcal{Q}_{W}^{-1}(\widehat{A})=A_{W}(\mathbf{x},\mathbf{p})=\int_{\mathbb{R}^{n}}\braket{\mathbf{x}-\frac{\mathbf{x}'}{2}|\widehat{A}|\mathbf{x}+\frac{\mathbf{x}'}{2}}e^{i\mathbf{p}\cdot\mathbf{x}'/\hbar}d\mathbf{x}',
\end{equation}
for any operator $\widehat{A}\in \mathrm{HS}(L^2(\mathbb{R}^n))$. Although these expressions were originally derived for square-integrable functions on phase space, they can be extended to arbitrary functions by employing generalized functions \cite{Reed}.

An associative and noncommutative product, commonly referred to as the star product, is induced in the algebra of classical observables $L^{2}(\mathbb{R}^{2n})$ as follows. Given that the Weyl's map $\mathcal{Q}_{W}:L^{2}(\mathbb{R}^{2n})\to {\rm HS}(L^{2}(\mathbb{R}^{n}))$ is bijective and the product of Hilbert-Schmidt operators is closed \cite{Takhtajan}, it implies that there is a unique function $f\star g\in L^{2}(\mathbb{R}^{2n})$, satisfying    
\begin{equation}
\mathcal{Q}_{W}(f)\mathcal{Q}_{W}(g)=\mathcal{Q}_{W}(f\star g)  \,.
\end{equation} 
The function $f\star g$, also called the Moyal star product is characterized by using the inverse Weyl's map as 
\begin{equation}\label{starWW}
(f\star g)(\mathbf{x},\mathbf{p})=\mathcal{Q}_{W}^{-1}(\mathcal{Q}_{W}(f)\mathcal{Q}_{W}(g)) \,.
\end{equation}
Thus, we may combine expressions (\ref{Weyl}) and (\ref{inverseWeyl}), to obtain
\begin{equation}\label{integralstar}
\hspace{-12ex}
(f\star g)(\mathbf{x},\mathbf{p})=\frac{1}{(\pi\hbar)^{2}}\int_{\mathbb{R}^{4n}}f(\mathbf{a},\mathbf{b})g(\mathbf{c},\mathbf{d})e^{-\frac{2i}{\hbar}(\mathbf{p}\cdot(\mathbf{a}-\mathbf{c})+\mathbf{x}\cdot(\mathbf{d}-\mathbf{b})+(\mathbf{c}\cdot\mathbf{b}-\mathbf{a}\cdot{d}))}d\mathbf{a}\,d\mathbf{b}\,d\mathbf{c}\,d\mathbf{d} \,,
\end{equation}  
which is known as the integral representation of the Moyal product. If we set $\hbar\to 0$, the above formula reduces to the product \cite{Folland}, 
\begin{equation}
\lim_{\hbar\to 0}f\star g=fg.
\end{equation}
In other words, this product is a deformation of the ordinary commutative pointwise product of functions on $\mathbb{R}^{2n}$. More generally, the Moyal product may be written as an asymptotic expansion in powers of the parameter $\hbar$, as detailed in \cite{Folland}.
In particular, for smooth functions defined on the phase space $f,g \in C^{\infty}(\mathbb{R}^{2n})$, with a Lie algebra structure induced by a Poisson bracket 
\begin{equation}
\left\lbrace f,g\right\rbrace=P(f,g)=\frac{\partial f}{\partial \mathbf{x}}\cdot\frac{\partial g}{\partial \mathbf{p}}-\frac{\partial f}{\partial \mathbf{p}}\cdot\frac{\partial g}{\partial \mathbf{x}}  \,,
\end{equation}
one can perform a Taylor expansion of both functions inside the integral (\ref{integralstar}), which leads to the Moyal star product in its differential representation,
\begin{equation}\label{Mproduct}
f\star g=fg+\sum_{n=1}^\infty \left(\frac{i\hbar}{2} \right)^{n} P^{n}(f,g) \,,
\end{equation}  
where the $\mathbb{R}$-bilineal map $P^{n}:C^{\infty}(\mathbb{R}^{2n})\times C^\infty(\mathbb{R}^{2n})$ corresponds to the $n$-th power of the Poisson bracket, interpreted as a bidifferential operator acting on $f$ and $g$ \cite{Bayen}. From the above discussion we observe that the bracket
\begin{equation}
\left\lbrace f,g \right\rbrace_{M}=\frac{1}{i\hbar}\left(f\star g-g\star f \right)  \,,  
\end{equation}
also known as the Moyal bracket, satisfies the properties of a Lie bracket. To be precise, it is skew-symmetric, bilinear and the Jacobi identity follows as a direct consequence of the associativity of the star product \cite{Kontsevich}. This implies that it is the Moyal bracket, rather than the Poisson bracket, that corresponds to the quantum commutator, and which reduces to the Poisson structure in the limit $\hbar \to 0$. Consequently, this observation underscores that quantum mechanics can be interpreted as a deformation of classical mechanics, with $\hbar$ playing the role of a deformation parameter. 

\section{Star exponentials and the Feynman-Kac formula}
\label{sec:FK}

The path integral quantization formalism remains a fundamental tool in the study of contemporary quantum mechanics and quantum field theory. In particular, this formalism has proven to be extremely effective in studying perturbative approximations across a wide range of physical phenomena \cite{Kleinhert}. In a pioneering work published in 1979, Pankaj Sharan established a connection between the star exponentials emerging within the deformation quantization formalism and Feynman’s path integrals \cite{Sharan}. Specifically, he demonstrated that Feynman's path integrals in quantum mechanics can be expressed as the Fourier transform of the star exponential associated with the Moyal product. These results were later extended to the context of field theory in the holomorphic representation \cite{Dito}, \cite{coherent} and, more recently, the star exponential of a Hamiltonian function has been determined in terms of quantum mechanical propagators \cite{starexpo}. In this section, our aim is to derive the Feynman-Kac formula by using the deformation quantization approach. This formula relates the asymptotic behavior of the time dependent propagators to the ground state energy level of a quantum system. For simplicity, we will focus on systems with a single degree of freedom, but the generalization to more dimensions follows straightforwardly. We conclude this section with a few illustrative physical examples.

In the framework of deformation quantization, the computation of the spectrum and the determination of the projectors associated with a given observable $H$ is achieved through the use of the star exponential, a particular function on phase space defined as
\begin{equation}\label{Exp}
\Exp\left(-\frac{i}{\hbar}tH \right)\equiv \sum_{n=0}^{\infty}\frac{1}{n!}\left(-\frac{i}{\hbar}t \right)^{n}H^{*n}   \,,  
\end{equation}  	
where $H^{*n}=H\star H\cdots\star H$ ($n$ factors). In the case where $H$ corresponds to the Hamiltonian of the system, the star exponential has a Fourier-Dirichlet expansion \cite{Bayen}, given by
\begin{equation}\label{FDexpansion}
\Exp\left(-\frac{i}{\hbar}tH \right)=2\pi\hbar\sum_{n=0}^{\infty}e^{-\frac{i}{\hbar}tE_{n}}\rho_{n}  \,,
\end{equation}
where $E_{n}$ denotes the eigenvalues of the Hamiltonian operator $\widehat{H}=\mathcal{Q}_{W}(H)$, and the phase space functions $\rho_{n}$ represent the diagonal Wigner functions. These Wigner functions are obtained by applying the inverse Weyl's quantization map (\ref{inverseWeyl}) to the projection operator $\widehat{P}_{\ket{n}}=\ket{n}\bra{n}$, associated with the normalized energy eigenstate $\ket{n}$. An overall factor $1/2\pi\hbar$ is included to guarantee that the Wigner function $\rho_{n}:=\frac{1}{2\pi\hbar}\mathcal{Q}^{-1}_{W}(\widehat{P}_{\ket{n}})$ turns out to be normalized in phase space
\begin{equation} \label{normalization}
\int_{\mathbb{R}^{2}}\rho_{n}(x,p)dx\,dp=1.
\end{equation} 
Furthermore, for a continuous spectrum, the summations in the previous expressions must be replaced by integrals over a continuous variable representing the energy \cite{Prata, Cosmas}.

The Feynman-Kac formula in the context of deformation quantization formalism follows from the Fourier-Dirichlet expansion (\ref{FDexpansion}). This expansion establishes a connection between the asymptotic behavior of the star exponential of the Hamiltonian as $t\to-i\infty$ and the ground state energy level a the quantum system. In order to see this, let us integrate the formula (\ref{FDexpansion}) over the entire phase space, we find that
\begin{equation}\label{sum}
\frac{1}{2\pi\hbar}\int_{\mathbb{R}^{2}}\Exp\left(-\frac{i}{\hbar}tH \right)\,dx\,dp=\sum_{n}^{\infty}e^{-\frac{i}{\hbar}tE_{n}},    	
\end{equation}
where we have made use of the normalization property of the Wigner functions (\ref{normalization}) for each value of $n$. By performing an analytic continuation of the time variable via a Wick rotation, $\tau=it$, the summation on the right hand side of equation (\ref{sum}) becomes dominated by the ground state, as the contributions from higher energy states decay exponentially faster. We obtain the asymptotic behavior
\begin{equation}
\lim_{\tau\to\infty}\frac{e^{\frac{\tau}{\hbar}E_{0}}}{2\pi\hbar}\int_{\mathbb{R}^{2}}\Exp\left(-\frac{\tau}{\hbar}H \right)\,dx\,dp=1,
\end{equation}    
where $E_{0}$ corresponds to the non-degenerate ground state energy. If the system has a degenerate ground state, the former integral will be equal to the degree of degeneracy. From this expression, the ground state energy can be extracted as
\begin{equation}\label{FK}
E_{0}=-\lim_{\tau\to\infty}\frac{\hbar}{\tau}\ln \left[\frac{1}{2\pi\hbar}\int_{\mathbb{R}^{2}}\Exp\left(-\frac{\tau}{\hbar}H \right)\,dx\,dp \right]. 
\end{equation}
This result constitutes the Feynman-Kac formula within the framework of deformation quantization. Remarkably, this formula provides an efficient method for determining the ground state energy levels without relying on the use of operators or Green's functions \cite{Glimm}. Instead, it solely depends on the asymptotic behavior of the star exponential of the Hamiltonian which is either a smooth function or a distribution defined on phase space.

\subsection{Examples}
In this subsection, we present some well-known examples with the main aim of explicitly illustrating the implementation  of the Feynman-Kac formula using the star exponential formalism.  


\subsubsection{Free particle.}
As a first test case, let us consider a free particle in one dimension. The Wick rotation of the star exponential corresponding to the free particle Hamiltonian, $H_{\mathrm{free}}(x,p)=p^{2}/2m$, reads \cite{starexpo}, \cite{Cosmas}, 
\begin{eqnarray}
\Exp\left(-\frac{\tau}{\hbar}H_{\mathrm{free}}(x,p)\right)
&=& e^{-\frac{\tau}{\hbar}\left(\frac{p^{2}}{2m}\right)} \, .
\end{eqnarray}
Since the star exponential is represented by a Gaussian function, we can make use of the Feynman-Kac formula (\ref{FK}) to obtain that the corresponding ground state energy for a free particle is given by $E_0=0$.

\subsubsection{Harmonic oscillator.}
Our second example is devoted to analyze the case of a quantum harmonic oscillator in one dimension, with Hamiltonian $H_{\mathrm{ho}}(x,p)=p^{2}/2m+m\omega^{2}x^{2}/2$. For this instance, the Wick-rotated star exponential reads \cite{starexpo},
\begin{eqnarray}\label{SEho}
\hspace{-5ex}\Exp\left(-\frac{\tau}{\hbar}H_{\mathrm{ho}}(x,p)\right)&=& \left(\cos\left( \frac{-i\omega\tau}{2}\right)  \right)^{-1}\exp\left[\frac{2H_{\mathrm{ho}}(x,p)}{i\omega\hbar}\tan\left(\frac{-i\omega \tau}{2} \right) \right].   
\end{eqnarray}
The integration of the star exponential over phase space yields\begin{equation}
  \frac{1}{2\pi\hbar}\int_{\mathbb{R}^{2}}\Exp\left(-\frac{\tau}{\hbar}H_{\mathrm{ho}}(x,p)\right)\,dx\,dp =\frac{1}{2}\csch\left(\frac{\omega \tau}{2}\right).
\end{equation}
By rewriting the hyperbolic cosecant as 
\begin{equation}
\frac{1}{2}\csch\left(\frac{\omega \tau}{2} \right)=\frac{e^{\omega\tau/2}}{e^{\omega\tau}-1},
\end{equation} 
the Feynman-Kac formula for the ground state energy becomes
\begin{eqnarray}
E_{0}&=&-\lim_{\tau\to\infty}\frac{\hbar}{\tau}\ln \left[\frac{e^{\omega\tau/2}}{e^{\omega\tau}-1}\right], \nonumber \\
&=&\frac{\hbar\omega}{2},
\end{eqnarray}
which corresponds to the minimal energy for the harmonic oscillator.

\subsubsection{Linear potential.}
Let us now consider the Hamiltonian $H_{\mathrm{lp}}(x,p)=p^{2}+x$, which corresponds to a linear potential in one dimension. In this case, after performing a Wick rotation, the star exponential is expressed as \cite{starexpo},
\begin{eqnarray}
\Exp\left(-\frac{\tau}{\hbar}H_{\mathrm{lp}}(x,p)\right)&=& e^{-\frac{\tau}{\hbar}(x+p^{2}-\tau^{2}/12)}  \,.
\end{eqnarray}
Due to the cubic behavior of the exponential in the imaginary time variable, the application of the Feynman-Kac formula reveals that, in this case, the energy is not bounded from below.	

\subsubsection{General quadratic form.}
Our next example is devoted to general quadratic Hamiltonians of the form  
\begin{equation}
    H_{\mathrm{q}}(x,p)=ap^{2}+bx^{2}+2cxp,
\end{equation}
where $a,b,c \in\mathbb{R}$. These Hamiltonians are particularly significant due to their invariance under transformations generated by the group $SL(2,\mathbb{R})$. In this case, after a Wick rotation, the star exponential takes the form,
\begin{equation}
    \Exp\left(-\frac{\tau}{\hbar}H_{\mathrm{q}}(x,p)\right)=\frac{1}{\cosh(\sqrt{ab-c^{2}}\tau)}e^{-\frac{iH_{\mathrm{q}}(x,p)}{\hbar\sqrt{ab-c^{2}}}\tanh{\sqrt{ab-c^{2}}\tau}}.
\end{equation}
  This star exponential function generates a double covering group of $SL(2,\mathbb{C})$, which is a simply connected group \cite{Omori}. To determine the ground state energy using the Feynman-Kac formula, we need to integrate over the entire phase space, as expressed by (\ref{FK}), thus yielding
  \begin{eqnarray}\label{vacuum}
      \frac{1}{2\pi\hbar}\int_{\mathbb{R}^{2}}\Exp\left(-\frac{\tau}{\hbar}H_{\mathrm{q}}(x,p)\right)\,dx\,dp =\frac{1}{2}\csch\left(\sqrt{ab-c^{2}} \tau\right).
  \end{eqnarray}
Then, 
\begin{eqnarray}
    E_{0}&=&-\lim_{\tau\to\infty}\frac{\hbar}{\tau}\ln \left[\frac{1}{2}\csch\left(\sqrt{ab-c^{2}} \tau\right)\right], \nonumber \\
&=&\hbar\sqrt{ab-c^{2}}.
\end{eqnarray}
Equilibrium states with energy $E_{0}$ can be directly obtained as limit points of the star exponential function (\ref{vacuum}), in accordance with the Kubo-Martin-Schwinger (KMS) condition \cite{Basart}.  

\subsubsection{Damped harmonic oscillator.} As our last example we consider a damped harmonic oscillator whose dynamics is governed by
\begin{equation}
    \ddot{x}+2\gamma \dot{x}+\omega^{2}x=0,
\end{equation}
where $\gamma>0$ denotes the damping factor and $\omega$ the frequency. In this example, dissipative effects can be incorporated into the algebra of observables by introducing a suitable deformation of the Moyal product, with $\gamma$ serving as the deformation parameter \cite{Damped}. More precisely, the star product is given by
\begin{equation}
f\star_{\gamma} g=fg+\sum_{n=1}^\infty \left(\frac{i\hbar}{2} \right)^{n} M^{n}(f,g) \,,
\end{equation} 
where 
\begin{equation}\label{Mbracket}
M(f, g) = \frac{\partial f}{\partial q} \frac{\partial g}{\partial p} - \frac{\partial f}{\partial p} \frac{\partial g}{\partial q} - 2\gamma m \frac{\partial f}{\partial p} \frac{\partial g}{\partial p}.
\end{equation}
This star product appears as a $\gamma$-deformation of the Moyal product (\ref{Mproduct}), in the sense that in the limit $\gamma\to 0$, we recover the standard Moyal product. The bracket $M(f,g)$ in Eqn. (\ref{Mbracket}), can be interpreted as a modified Poisson bracket incorporating a Hochschild differential term proportional to $\gamma$. This modification implies that the star product $\star_{\gamma}$ is mathematically equivalent to the Moyal product, as discussed in \cite{Bayen}. Therefore, the star exponential for the damped harmonic oscillator reads
\begin{eqnarray}
  \hspace{-13ex}   \Expg\left(-\frac{it}{\hbar}H_{\mathrm{d}}(x,p)\right)&=&\frac{\exp(\gamma t/2)}{\cos(\omega t/2)(1+\frac{2\gamma}{\omega}\tan(\omega t/2))} \nonumber \\
     && \times \exp\left[ \frac{-i}{\hbar\omega}\tan(\omega t/2)\left(m\omega^{2}x^{2}+\frac{p^{2}}{m(1+\frac{2\gamma}{\omega}\tan(\omega t/2))}\right)\right]. 
\end{eqnarray}
After applying a Wick rotation, $\tau=it$, and integrating over the entire phase space, we obtain 
\begin{equation}
     \frac{1}{2\pi\hbar}\int_{\mathbb{R}^{2}}\Expg\left(-\frac{\tau}{\hbar}H_{\mathrm{d}}(x,p)\right)\,dx\,dp =\frac{1}{2}e^{-i\gamma\tau/2}\csch\left(\omega\tau/2 \right).
\end{equation}
Then,
\begin{eqnarray}
     E_{0}&=&-\lim_{\tau\to\infty}\frac{\hbar}{\tau}\ln \left[\frac{1}{2}e^{-i\gamma\tau/2}\csch\left(\omega\tau/2 \right)\right] \nonumber \\
&=& \hbar\omega(1/2+i\gamma/2\omega),
\end{eqnarray}
which corresponds to the ground state energy of the damped harmonic oscillator. Notice that the spectrum is now complex, reflecting the fact that we are dealing with a non-Hermitian product or, equivalently, that the corresponding Hamiltonian is not a self-adjoint operator \cite{Damped}. 
\section{Conclusions}
\label{sec:conclu}

In the present work a relation between the star exponential function of deformation quantization and Feynman-Kac's formula from the path integral approach of quantum mechanics was explored. The Feynman-Kac formula is usually used to determine the ground state energy of the system. In this work after a short overview of the deformation quantization formalism in Weyl's(-Wigner) formalism, it is applied to establish an explicit link between the star exponential and the Feynman-Kac formula. In order to do that,  we have considered the star exponential of the Hamiltonian of a given mechanical system as argument. Upon a Wick rotation it is shown that the ground state energy of the system is related to the asymptotic limit of the phase space integration of this star exponential of the Hamiltonian. The obtained formula is written in Eq. (\ref{FK}). Moreover, in Section 3 some examples for different physical systems are given. In all those examples the ground state energy  was obtained in an efficient way compared with the usual techniques described in \cite{Glimm}. 
 
In the process we have established a new contact point between the deformation quantization formalism via the star exponential and the path integral approach of quantum mechanics. In a previous work by the authors \cite{starexpo} it was found also a contact point between the star exponential of a physical system described by a Hamiltonian and its corresponding propagator in the path integral approach. Thus, the results obtained in the present article contribute to establish a clearer correspondence between the two approaches to Quantum Mechanics. Further, it is clear and interesting that these results would be straightforwardly generalized to the quantum field and string theories. Some attempts in this direction will be reported in a future communication.

\section*{Acknowledgments}
The authors would like to acknowledge support from SNII CONAHCYT-Mexico. JBM acknowledge financial support from Marcos Moshinsky foundation and thanks to the Dipartimento di Fisica ``Ettore Pancini" for the kind invitation and its generous hospitality.  AM acknowledges financial support from COPOCYT under project 2467 HCDC/2024/SE-02/16 (Convocatoria 2024-03, Fideicomiso 23871).

\section*{References}

\bibliographystyle{unsrt}

\end{document}